\documentclass[journal=jacs,linenumbers,manuscript=letter]{achemso}
\usepackage{graphicx}
\usepackage{amsmath, amsthm}
\usepackage{amssymb}
\usepackage[labelfont=bf,labelsep=space]{caption}
\usepackage{epstopdf}
\usepackage{threeparttable}
 \usepackage{lineno}
 \usepackage{tabularx}

\title{Polarization Dependence of Water Adsorption to CH$_{3}$NH$_{3}$PbI$_{3}$ (001) Surfaces}
\author{Nathan Z. Koocher, Diomedes Saldana-Greco, Fenggong Wang, Shi Liu, and Andrew M. Rappe}
\affiliation{The Makineni Theoretical Laboratories, Department of Chemistry, University of Pennsylvania, Philadelphia, PA 19104--6323 USA}
\email{rappe@sas.upenn.edu}

\begin{document}
\maketitle

\begin{abstract}
The instability of organometal halide perovskites when in contact with water is a serious challenge to their feasibility as solar cell materials.  Although studies of moisture exposure have been conducted, an atomistic understanding of the degradation mechanism is required.  Toward this goal, we study the interaction of water with the (001) surfaces of CH$_{3}$NH$_{3}$PbI$_{3}$ under both low and high water concentrations using density functional theory.  We find that water adsorption is heavily influenced by the orientation of the methylammonium cations close to the surface.  It is demonstrated that, depending on methylammonium orientation, the water molecule can infiltrate into the hollow site of the surface and get trapped.  Controlling dipole orientation via poling or interfacial engineering could thus enhance its moisture stability.  We do not see a direct reaction between the water and methylammonium molecules.  Furthermore, calculations with an implicit solvation model indicate that a higher water concentration may facilitate degradation.   
\end{abstract}
\newpage

%

\section*{\label{sec:level1} Introduction}

Solar cells based on organometal halide perovskite (OMHP), especially methylammonium lead iodide (MAPbI$_{3}$), have had a remarkable increase in efficiency in the past five years~\cite{Zhou14p542,Jeon14p7837,KRICT}.  Although the photovoltaic mechanism underlying this high power conversion efficiency is not fully understood, the impressive efficiency of MAPbI$_{3}$ is related to its suitable band gap, good carrier transport properties, high optical absorption, and long diffusion length~\cite{Papavassiliou95p1713,Noel14px,Umari14p4467,Chiarella08p045129,
Ogomi14p1004,Eperon14p982,Stoumpos13p9019,Mosconi14p16137,Filip14p245145}.  Despite its promising efficiency, commercial use of the OMHP-based solar cell is limited in part by its poor stability with respect to moisture.~\cite{Bass14p15819, Hailegnaw15p1543} One proposed work-around involves packaging the OMHP with hydrophobic materials to enhance stability~\cite{Jung14p10, Hwang15pAccepted}, but this has considerable drawbacks including more complicated device architecture, additional manufacturing requirements, interfacial defects, and necessitating insight into interfacial atomic and electronic structures.~\cite{Eperon14p151, Shi15p2472, Roiati14p2168, Mosconi14p2619, Yang15p4229, Yin15p1396, Lindblad14p648}  Understanding the mechanism of degradation is therefore critical to providing materials design principles and engineering strategies for achieving long-term stability.  

The degradation mechanism is currently an open question. Niu and coworkers proposed that methylammonium iodide (MAI) and PbI$_{2}$ are first formed, with further breakdown of MAI into methylamine (CH$_{3}$NH$_{2}$) and hydrogen iodide (HI), concluding with the formation of I$_{2}$(solid) and H$_{2}$(gas) after exposure to oxygen and sunlight~\cite{Niu14p705,Niu15p8970}. Mechanistically, Frost and coworkers suggested that water abstracts a hydrogen from the MA molecule in an acid-base reaction which leads to the formation of HI, CH$_{3}$NH$_{2}$, and PbI$_{2}$~\cite{Frost14p2584}. Further investigation of the initial steps of the degradation process have been performed.  Leguy and colleagues suggested that the MAPbI$_{3}$ responds differently to moisture depending on the water concentration based on experimental observations.~\cite{Leguy15p3397} When exposed to low humidity, MAPbI$_3$ forms a transparent monohydrate (MAPbI$_{3}$$\cdot$H$_{2}$O), which can be dehydrated back to MAPbI$_{3}$ by raising the temperature. After prolonged exposure to water vapor, the monohydrate converts to a dihydrate ((MA)$_{4}$PbI$_{6}$$\cdot$2H$_{2}$O), which eventually dissolves in water, leading to decomposition.\cite{Leguy15p3397,Hailegnaw15p1543, Imler15p11290} It is also suggested that water could penetrate into the perovskite along grain boundaries and that irreversible decomposition occurs when a grain boundary has completely converted to the monohydrate.~\cite{Leguy15p3397}  The proposed hydration of grain boundaries is not fully understood and will benefit from an atomistic examination of how water penetrates into the perovskite. Both Christians \emph{et al}.\  and Yang \emph{et al}.\ found hydrate formation, although the exact nature of the hydrate species is ambiguous.~\cite{Christians15p1530,Yang15p1955}  Both groups observed at least partial recovery of the perovskite when the system was dehydrated. Moreover, Christians \emph{et al}.\ found that the perovskite decomposes differently in dark and light conditions;~\cite{Christians15p1530} the presence of moisture and illumination cause the perovskite to degrade to PbI$_{2}$, but in the dark, PbI$_{2}$ is not formed.  It should be noted that Philippe \emph{et al}.\ do not see evidence of a hydrate state in their photoelectron spectroscopy data, but attribute it to the instability of the hydrate phase.~\cite{Philippe15p1720}  Although experiments are providing an increasingly clear picture of the degradation mechanism, an atomistic-level understanding of the process could uncover new methods to stabilize the material. 

The observation of a monohydrate state suggests that the degradation mechanism starts with a surface-environment interaction process. Theoretical surface studies of this system so far have investigated stable terminations of tetragonal~\cite{Haruyama14p2903} and orthorhombic~\cite{Wang15p1136} MAPbI$_{3}$, as well as adsorption of anisole (a hole-transport material proxy) on (001) surfaces of pseudocubic MAPbI$_{3}$~\cite{Torres14p26947}, and the interaction of water with these surfaces with first principles molecular dynamics~\cite{Mosconi15p4885} and density functional theory.~\cite{Tong15p0}  In this paper, we present an atomistic perspective of water interacting with the relevant PbI$_{2}$- and MAI-terminated (001) surfaces of MAPbI$_{3}$ possessing different polarity. To simulate low relative humidity conditions, we introduce one explicit water molecule on the surface per primitive surface cell, and to simulate high relative humidity conditions, we use an implicit solvation model with and without an explicit water molecule. We show that surfaces with different terminations and polarities respond differently to water, leading to different surface bonding configurations.~\cite{Kolpak07p166101, Li08p473}   We also investigate the effect of water penetration into the material. We find that the MA dipole orientation strongly affects the surface-water interaction at low coverages and postulate that control of the orientation through poling with an electric field or interfacial engineering between the perovskite material and capping layers could enhance the moisture stability of the material.  Also, the implicit solvation model results in the elongation of the lattice vector perpendicular to the surface, suggesting that the dielectric response of a higher water concentration aids in material degradation by expanding the lattice.  

The polarization of a material is known to affect the surface adsorption of small molecules.~\cite{Li08p473,Spanier06p735,Fong06p127601,Kolpak07p166101,Wang09p047601,Garra09p1106} Due to the permanent dipole moment of the MA cation, there are two orientations that produce polarization extrema: all molecules either having their NH$_{3}$$^{+}$-ends pointing toward the surface, termed \emph{P}$^{+}$, or all molecules have their CH$_{3}$-ends pointing toward the surface, termed \emph{P}$^{-}$. The top-down and side views of the bare surfaces considered are shown in Fig.\ 1.

\section*{\label{sec:level1} Methodology}
The surfaces were modeled using a slab model within the Quantum ESPRESSO DFT package.\cite{Giannozzi09p395502}  The PBE-GGA functional~\cite{Perdew96p3865}  was used as well as norm-conserving, optimized pseudopotentials with a plane-wave cutoff of 50 Ry, generated with the OPIUM code.~\cite{Rappe90p1227,Ramer99p12471}  The slab model for both the MAI- and PbI$_{2}$-terminated surfaces had 9 layers, including the water adsorbate, and a vacuum of about 15 \AA\ between each slab. Geometry optimization calculations involved relaxing the top four layers and the adsorbates, until the forces on atoms were less than 0.01 eV/\AA\ in all directions, while the bottom five layers were fixed to the tetragonal bulk structure.  A $4\times4\times1$ Monkhorst-Pack $k$-point grid was used for the relaxation, while a finer grid was used as needed for the density of states calculations. A linear polarization continuum model of water implemented in jDFTx~\cite{Petrosyan05p15436,Gunceler13p074005} was used to obtain surface structures under liquid water environment.

\begin{figure}[H]
\centering
\includegraphics[width=0.6\textwidth, angle = 90]{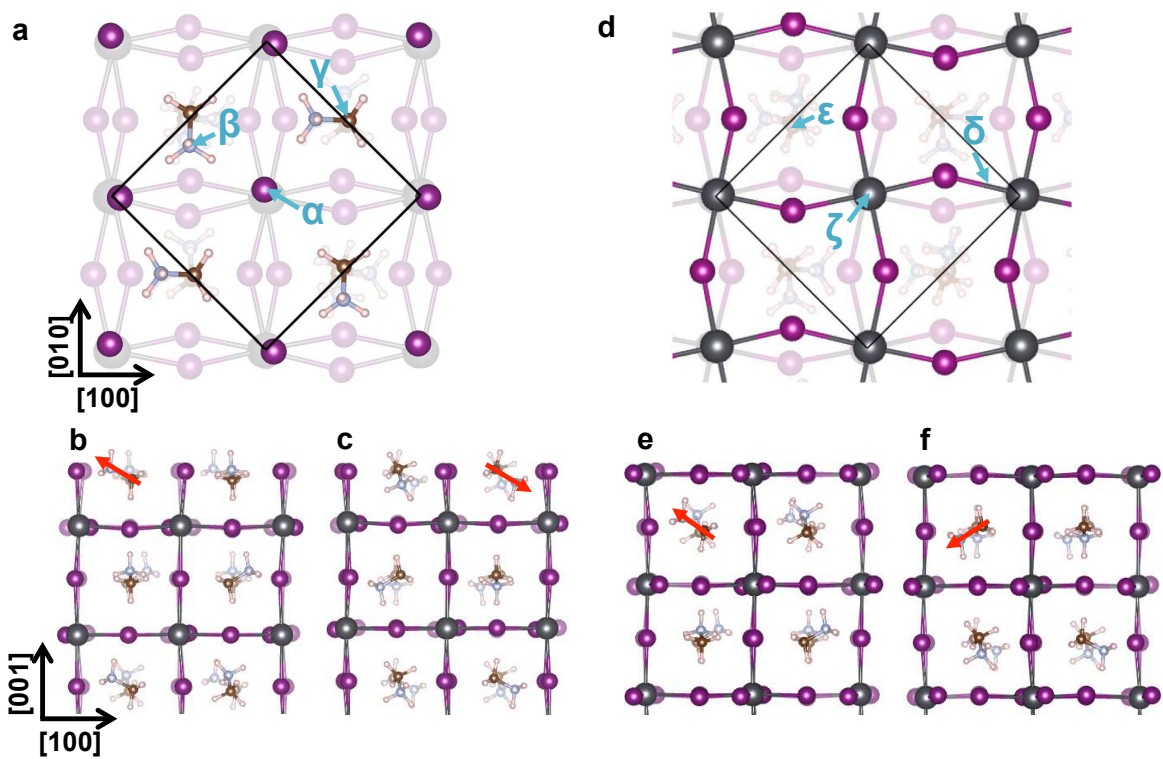}\\
\label{}
 \caption{\textbf{Water adsorption sites on the MAPbI$_{3}$ (001) surface.} \textbf{(a)} Top-down view of the MAI-termination, showing exposed MA molecules and associated iodine atoms. Red letters indicate studied water adsorption sites, $\alpha$$^{-|+}$, $\beta$$^{-|+}$, and $\gamma$$^{-|+}$.  \textbf{(b)} Side view of the MAI-terminated \emph{P}$^{+}$ surface, where the NH$_{3}$$^{+}$-end is exposed. \textbf{(c)} Side view of the MAI-terminated \emph{P}$^{-}$ surface, where the CH$_{3}$-end of the MA molecule is exposed. \textbf{(d)} Top-down view of the PbI$_{2}$-termination, showing coordinated Pb atoms with associated iodine atoms; both are labeled to facilitate discussion. Red letters indicate water adsorption sites, $\delta$$^{-|+}$, $\epsilon$$^{-|+}$, and $\zeta$$^{-|+}$.   \textbf{(e)} Side view of the PbI$_{2}$-terminated \emph{P}$^{+}$ surface.  The NH$_{3}$$^{+}$-end of the subsurface MA molecule is pointed toward the surface. \textbf{(f)} Side view of the PbI$_{2}$-terminated \emph{P}$^{-}$ surface. The CH$_{3}$-end of the subsurface MA molecule is pointed toward the surface. The dipole of the MA molecule in each slab is indicated with a red arrow. The black lines represent the surface periodicity studied ($\sqrt{2}$$\times$$\sqrt{2}$ $R$ 45$^{\circ}$). Grey: Pb, Purple: I, Blue: N, Brown: C, White: H.}
 \end{figure}
 
 \begin{table}
\caption{Adsorption energies, $E$$_{\rm ads}$ in eV, of water on each of the studied sites of MAI- and PbI$_{2}$-terminated surfaces, with the MA dipoles aligned in the $P$$^{-}$ or $P$$^{+}$ arrangement.  Note that on the PbI$_{2}$-terminated \emph{P}$^{-}$ surface, water molecules starting at $\delta$$^{-}$ and $\epsilon$$^{-}$ relax to $\zeta$$^{-}$.  Explanation for this is given in Figure 2.}
\begin{tabular}{|l|l|l|l|l|l|l|l|}
\hline
\multicolumn{4}{|l|}{MAI-termination}                          & \multicolumn{4}{l|}{PbI$_{2}$-termination}                         \\ \hline
\multicolumn{2}{|l|}{\emph{P}$^{+}$} & \multicolumn{2}{l|}{\emph{P}$^{-}$} & \multicolumn{2}{l|}{\emph{P}$^{+}$} & \multicolumn{2}{l|}{\emph{P}$^{-}$} \\ \hline
site                       & \emph{E}$_{\rm ads}$ (eV)     & site                            &  \emph{E}$_{\rm ads}$ (eV)    & site                        & \emph{E}$_{ads}$ (eV)     & site                         & \emph{E}$_{\rm ads}$ (eV)    \\ \hline
$\alpha$$^{+}$      & -0.49                                 & $\alpha$$^{-}$          & -0.18                                 & $\delta$$^{+}$        & -0.40                                 & $\delta$$^{-}$        & $\rightarrow$  $\zeta$$^{-}$       \\ \hline
$\beta$$^{+}$        & -0.36                                 & $\beta$$^{-}$            & -0.13                                & $\epsilon$$^{+}$     & -0.38                                 & $\epsilon$$^{-}$     & $\rightarrow$  $\zeta$$^{-}$        \\ \hline
$\gamma$$^{+}$   & -0.48                                 & $\gamma$$^{-}$       & -0.28                                & $\zeta$$^{+}$          & -0.39                                 & $\zeta$$^{-}$         & -0.54                                  \\ \hline
\end{tabular}
\end{table}

\section*{\label{sec:level1} Results and Discussion}
We explore different adsorption sites for water on the surfaces by putting the molecule at the sites labelled in Fig.\ 1 and allowing the system to fully relax.  As we will discuss later, in some cases, different adsorption sites may lead to the same final structure after relaxation.  Sites on the MAI-terminated surfaces include: on top of an iodine atom, site $\alpha$; above a hydrogen of the exposed group of the MA molecule (ammonium, NH$_{3}$$^{+}$ for $P^{+}$; methyl, CH$_{3}$ for $P^{-}$), site $\beta$; and above a hydrogen of the lower group of the MA molecule (CH$_{3}$ for $P^{+}$, NH$_{3}$$^{+}$ for $P^{-}$), site $\gamma$. The corresponding sites on the $P^{-}$ and $P^{+}$ surfaces are denoted with $^{-}$ and $^{+}$, respectively and are termed $\alpha$$^{-|+}$, $\beta$$^{-|+}$, and $\gamma$$^{-|+}$. Sites on the PbI$_{2}$-terminated $P^{-}$ surface include: along the Pb-I bond with oxygen pointed toward Pb and H pointed toward I, site $\delta$$^{-}$; above the hollow site of the surface, site $\epsilon$$^{-}$; and above the Pb, sites $\zeta$$^{-}$. The corresponding sites on the $P^{+}$ surface are $\delta$$^{+}$, $\epsilon$$^{+}$, and $\zeta$$^{+}$, respectively.  The adsorption energies of water at each site after relaxation are evaluated as $E$$_{\rm{ads}}$\ =\ $E$$_{\rm{surface-H_{2}O}}$\ -- \ [$E$$_{\rm{bare-surface}}$\ +\ $E$$_{\rm{H_{2}O}}$], and are shown in Table 1. Water adsorption is favorable on all surfaces, although distinctly different interactions are found depending on both surface terminations and polarities. Adsorption to the MAI-terminated $P^{+}$ surface is more energetically favorable than to the corresponding \emph{P}$^{-}$ surface. For PbI$_{2}$-terminated surfaces, however, the water molecule prefers to bind to the \emph{P}$^{-}$ surface rather than to the \emph{P}$^{+}$ surface. This behavior is likely due to the competing hydrogen bond interactions between the MA molecules, the PbI$_{2}$ inorganic lattice, and the water molecules. For the MAI-terminated surface, water prefers to bind to the \emph{P}$^{+}$ surface with exposed NH$_{3}$$^{+}$ groups, as a hydrogen bond is formed between water and NH$_{3}$$^{+}$ groups, while no hydrogen bond is formed between water and CH$_{3}$ groups ($P^-$ surface). On the PbI$_{2}$-terminated \emph{P}$^{-}$ surface there is only weak interaction between the subsurface methyl groups and the surface PbI$_{2}$ lattice, so the interaction between water and the surface Pb is stronger than that on the \emph{P}$^{+}$ surface, where the NH$_{3}$$^{+}$ groups hydrogen bond to the inorganic lattice.~\cite{Lee15p6434}  Because all starting sites went to the same local minimum, we produced a contour plot which shows \emph{E}$_{\rm ads}$ as a function of position in the (001) plane, seen in Figure 2, indicating that there is only one local minimum around the Pb atom.  The atomic and electronic structure of the system resulting from various adsorption events will now be described.  

\begin{figure}
\centering
\includegraphics[width=0.8\textwidth]{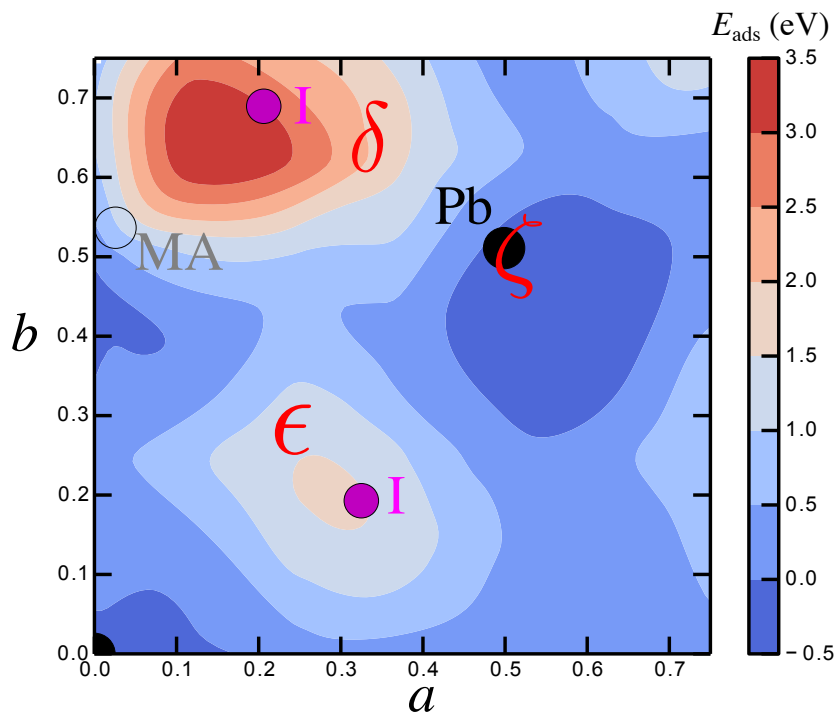}\\
\label{}
\caption{\textbf{Contour plot of water adsorption on the PbI$_{2}$-terminated \emph{P}$^{-}$ surface of MAPbI$_{3}$.}  Only one local minimum is seen at the Pb atom, explaining why all sampled structures in Table 1 resulted in the same adsorption energy.}
\end{figure}

Starting with the MAI-terminated surfaces, the lowest-energy configuration of water on the \emph{P}$^{+}$ surface is found from initial site $\alpha$$^{+}$ shown in Fig.\ 3a (side view).  The water is almost flat in the (110) plane with a hydrogen of water (H$_{\rm W}$) interacting with a surface iodine, but is slightly tilted to make a hydrogen bond between the oxygen of water (O) and one hydrogen atom (H$_{\rm N}$) from the MA NH$_{3}$$^{+}$ group.  To examine the bonding character in more detail, we compute the orbital projected density of states (PDOS), seen in Supplementary Information Figure S1.  It shows the overlap between the O \emph{p$_{y}$} orbital and the \emph{s} orbital of H$_{\rm N}$, indicating a hydrogen bond interaction.  On the MAI-terminated \emph{P}$^{-}$ surface, however, the orientation of the water molecule (as shown in Fig.\ 3b) is very different from that on the \emph{P}$^{+}$ surface.  In this case, relaxation with a water molecule initially at adsorption site $\alpha$$^{-}$ yields a structure where O and one H$_{\rm W}$ are located directly above the surface iodine, with the H$_{\rm W}$ 2.79 \AA\ above the iodine vertically. The PDOS of this structure, displayed in Figure S1, shows that the H$_{\rm W}$ \emph{s} orbital has weak overlap with the I \emph{p} orbital and that the O \emph{p} orbitals are split.  Comparing the PDOS highlights the weaker bonding of water to the \emph{P}$^{-}$ surface.  The water-\emph{P}$^{+}$ surface configuration is 0.31 eV lower in energy than the water-\emph{P}$^{-}$ surface configuration, probably due to the strong electrostatic repulsion between the oxygen and iodine on the \emph{P}$^{-}$ surface without the compensation of the hydrogen bond. 

\begin{figure}
\centering
\includegraphics[width=0.8\textwidth]{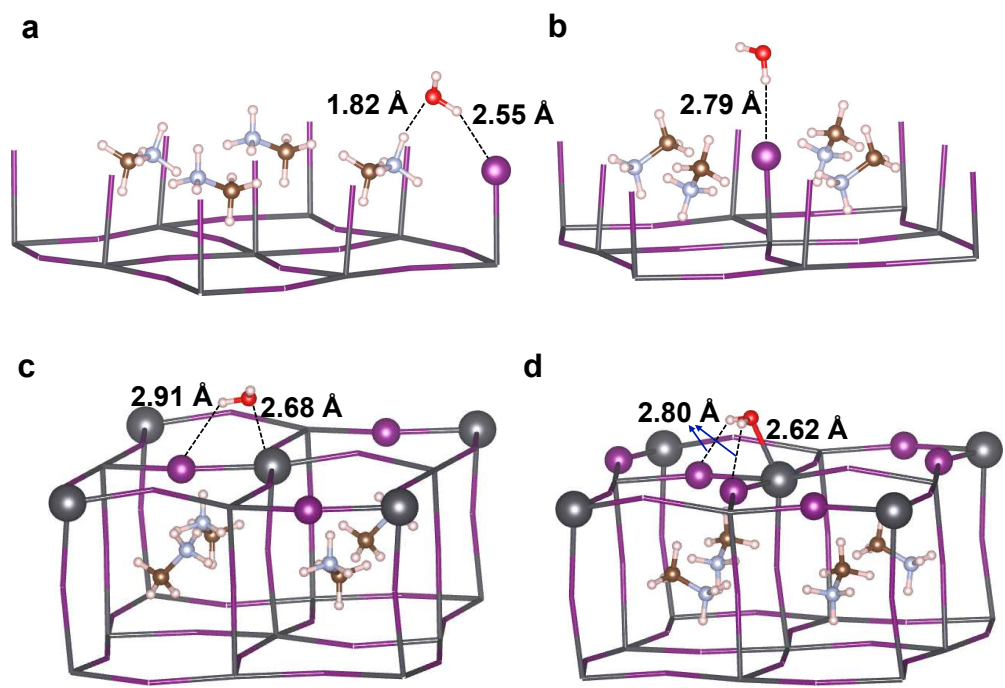}\\
\label{}
\caption{\textbf{Water adsorption to the MAI- and PbI$_{2}$-terminated surfaces with different polarities.}  Lowest-energy structure of water on the \textbf{(a)} MAI-terminated \emph{P}$^{+}$ surface, \textbf{(b)} MAI-terminated \emph{P}$^{-}$ surface, \textbf{(c)} PbI$_{2}$-terminated \emph{P}$^{+}$ surface, and \textbf{(d)} PbI$_{2}$-terminated \emph{P}$^{-}$ surface.} 
\end{figure}

These results suggest that the MAI-terminated \emph{P}$^{-}$ surface is more water resistant than the \emph{P}$^{+}$ surface due to the lack of water reactivity with the CH$_{3}$-end of the MA molecule.  Thus, exposing the CH$_{3}$  to water by applying an electric field~\cite{Chen15p7699} or by inducing local, interfacial ordering of the dipoles with a capping layer~\cite{Roiati14p2168} could improve stability.  If, however, the water molecule is approximately 2.00 \AA\ from  the NH$_{3}$$^{+}$-end of the MA molecule of the \emph{P}$^{-}$ surface (adsorption site $\gamma$$^{-}$), the MA dipole locally orients to favor the NH$_{3}$$^{+}$ hydrogen bonding with the water, leading to a 0.093 eV lower-energy configuration of water on the \emph{P}$^{-}$ surface.  Based on previous molecular dynamics studies, water travelling from vacuum to this distance is likely spontaneous.~\cite{Mosconi15p4885, Tong15p0}  This indicates that the surface is not water resistant in all cases, implying that an electric field might have to be applied continuously in the presence of water to improve stability.  In general, we see no evidence of chemical reactivity between the H$_{2}$O and the MA molecule, leaving the molecular cations intact regardless of orientation or proximity.  This is consistent with first principles molecular dynamics results showing that degradation is a structural rather than acid-base effect.~\cite{Mosconi15p4885}  We cannot rule out an acid-base mechanism, however, because we do not sample the full pH range in our calculation.~\cite{Frost14p2584,Yang15p1955,Christians15p1530,Leguy15p3397}

The lowest-energy water adsorbate configuration on the PbI$_{2}$-terminated \emph{P}$^{+}$ surface is shown in Fig.\ 3c. This structure is obtained from the relaxation of the water molecule initially on site $\delta$$^{-}$ (along a Pb-I bond with the oxygen of water pointing toward the Pb).  The water molecule becomes almost planar in the (001) plane, with the O making a 2.68 \AA\ bond with the Pb, while one hydrogen from the water interacts with a surface iodine 2.91~\AA\ away, as indicated in Fig.\ 3c.  On the PbI$_{2}$-terminated \emph{P}$^{-}$ surface, the water molecule starting on adsorption site $\delta$$^{+}$ ends up in a similar configuration as that on the \emph{P}$^{+}$ surface, but is closer to the surface, as shown in Fig.\ 3d. The water is oriented almost flat in the (001) plane above the surface, with the H$_{\rm W}$ atoms pointing slightly downward for better bonding with the surface iodine atoms. The O makes a 2.62 \AA\ bond with the surface Pb, while the H$_{\rm W}$ atoms interact with iodine atoms about 2.80 \AA\ away. Interestingly, these distances are shorter than in the case of water on the corresponding \emph{P}$^{+}$ surface site, indicating stronger bonding.  This message is reinforced by the splitting of the Pb \emph{p} orbitals in the \emph{P}$^{-}$ case, with no such splitting in the \emph{P}$^{+}$ case, as shown in Figure S1.  The change in bonding character is also reflected in the 0.14 eV larger adsorption energy of water on the \emph{P}$^{-}$ surface than on the \emph{P}$^{+}$ surface (Table 1). 

\begin{figure}
\centering
\includegraphics[width=0.8\textwidth]{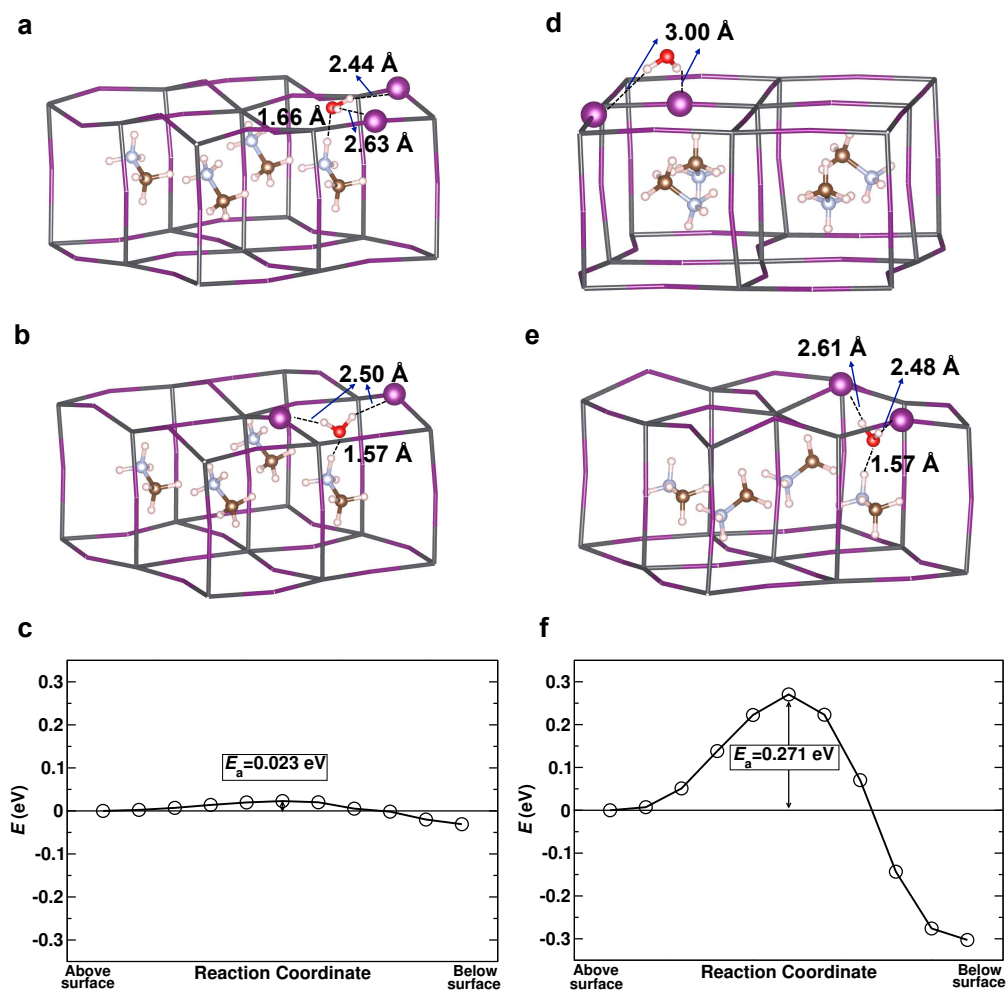}\\
\label{}
\caption{\textbf{Energetic and structural description of possible water penetration mechanism on the PbI$_{2}$-terminated surfaces with different polarities.} \textbf{(a)} Lowest-energy structure of water at the hollow site of the PbI$_{2}$-terminated \emph{P}$^{+}$ surface.
\textbf{(b)} Lowest-energy structure of water under the first layer of the PbI$_{2}$-terminated \emph{P}$^{+}$ surface.
\textbf{(c)} Reaction path of water traveling from the configuration seen in \textbf{(a)} to that in \textbf{(b)}.  The forward activation barrier is 0.023 eV, which can be overcoming using thermal energy.  This emphasizes the facility of water penetration on this surface.  
\textbf{(d)} Lowest-energy structure of water initially placed at the hollow site of the PbI$_{2}$-terminated \emph{P}$^{-}$ surface.  The water was repelled from the hollow site into the vacuum above the hollow site during relaxation.  
\textbf{(e)} Lowest-energy structure of water under the first layer of the PbI$_{2}$-terminated \emph{P}$^{-}$ surface, demonstrating that if water can penetrate the top layer, it will tilt the methylammonium dipole direction to form a hydrogen bond and become trapped.
\textbf{(f)} Reaction path of water traveling from the configuration seen in \textbf{(d)} to that in \textbf{(e)}.  The forward activation barrier is 0.27 eV, which is an order of magnitude higher than on the surface of opposite polarity.  This emphasizes the water-repelling nature of the surface and suggests paths of material stabilization based on poling with an electric field or interfacial engineering to create an ordered cation domain.}
\end{figure}

Experimental observations indicate that water molecules penetrate into the material along grain boundaries, converting the structure to the monohydrate state.~\cite{Leguy15p3397} Therefore, we explore the possibility of water penetrating through the hollow site in the PbI$_{2}$-terminated surfaces.  We model two scenarios on both \emph{P}$^{+}$ and \emph{P}$^{-}$ surfaces: H$_{2}$O in the plane of the first surface layer, and H$_{2}$O inside the material (under the first surface layer).  Note that this is different from adsorption site $\epsilon$ in Fig.\ 1 because $\epsilon$ corresponds to water in vacuum above the hollow site.  On the PbI$_{2}$-terminated \emph{P}$^{+}$ surface, the water that starts in the plane of the surface stays in-plane, with the O interacting via hydrogen bonding with the H$_{\rm N}$ of the subsurface MA molecule, while the two H$_{\rm W}$ atoms interact with surface iodine ions. This is shown in Figure 4a.  The PDOS of this structure displayed in Figure S2 shows that the O \emph{p$_{x}$} orbital and the H$_{\rm N}$ \emph{s}-orbital has appreciable overlap. This indicates that there is a moderate interaction between the two atoms, weakening octahedral rotations. The Pb-I-Pb bond angles are closer to 180$^{\circ}$, leading to a larger hollow site. Comparing the top-down view of the bare surface (Fig.\ S3) with that of this structure, it is clear that H$_{2}$O infiltration increases the size of a hollow site, and that the rotation suppression opens up the adjacent hollow sites, suggesting that water adsorption on the subsurface has a collective effect on the neighboring structure.

Manually moving the water through the hollow site of the \emph{P}$^{+}$ surface, such that the water gradually approaches the subsurface, yields a trapped, intact molecule (shown in Fig.\ 4b).  The PDOS for this structure (Figure S2) shows that the O \emph{p} orbitals are split, with the \emph{p$_{z}$} orbital having appreciable overlap with the $s$ orbital of H$_{\rm N}$ atom in the MA molecule, signaling strong hydrogen bonding.  To ensure that the structures in Figs. 4a and 4b are distinct minima, we performed a Nudged Elastic Band (NEB) calculation with those structures as endpoints in the path, shown in Fig.\ 4c.  We found that there is a 0.02 eV activation barrier going from configuration in Fig.\ 4a to that in Fig.\ 4b, while there is a 0.05 eV barrier for the reverse direction.  These barriers suggest that water could become trapped beneath the surface layer, with a facile equilibrium at room $T$.

The water was also placed in the \emph{P}$^{-}$ surface plane at the hollow site.  As expected based on the previously observed hydrophobicity of the methyl group, instead of staying in-plane, the water was repelled from the surface.  As seen in Fig.\ 4d, the optimized water orientation has the O pointed away from the surface and the two H$_{\rm W}$ atoms interacting with the surface iodine (Figure S2).  When the water is between the first and second layers in the PbI$_{2}$-terminated \emph{P}$^{-}$ system, the water molecule is trapped in part due to hydrogen bond formation (Figure S2) with the NH$_{3}$$^{+}$-end of the methylammonium, which is achieved through a change in orientation of the methylammonium cation.  We also calculated the activation energy barrier using NEB for this pair of structures (Figs. 4d and 4e) and found that the forward barrier is 0.27 eV and the reverse barrier is 0.57 eV, seen in Fig.\ 4f.  As expected, the forward barrier for this surface is higher than that for the corresponding \emph{P}$^{+}$ system and once the molecule penetrates the hollow site, it cannot leave.  Interestingly, the forward barrier is more than ten times the thermal energy, further indicating that the methyl end of the methylammonium can repel water.

Contrasting these surfaces, we see that the oxygen binds more strongly with the lead on the \emph{P}$^{-}$ surface than on the \emph{P}$^{+}$ surface and interacts more favorably with the ammonium end of the MA molecule than with the methyl end regardless of the surface termination. On the \emph{P}$^{-}$ surface, the water molecule is repelled from the hollow site, while on the \emph{P}$^{+}$ surface, the water molecule is stable in-plane with the top atomic layer, probably due to the strong hydrogen bond between the water and the NH$_{3}$$^{+}$ of the MA molecule.  Furthermore, both surfaces can trap water molecules in the subsurface, and such trapping disrupts the Pb-I-Pb angles of the surface layer.  These results suggest a possible scheme of how water interacts with the surface and possible methods to stabilize the material.  On approach to the surface, the water would first get physisorbed in a shallow energy well, forming a hydrate.  Then, environmental factors such as temperature and humidity could enable the H$_{2}$O to move to the surface, where it could get trapped at the interfacial plane of the hollow site on the \emph{P}$^{+}$ surface~\cite{Mosconi15p4885}, but still be repelled from the hollow site by the \emph{P}$^{-}$ surface.  If the water molecule penetrates the hollow site of the surface (of both polaritities), water and the PbI$_{2}$ inorganic lattice compete to form a hydrogen bond with MA, weakening the stability of MAPbI$_{3}$.  Most importantly, this suggests that the stability of MAPbI$_{3}$-solar cells could be enhanced if they were poled using an electric field or locally ordered via an interfacial substrate such that the methyl ends of the MA molecules were pointed toward the surface.  While our kinetic barrier results demonstrate that water incursion into the material is slowed, without knowing other steps leading to chemical reaction, it is difficult to assess the actual extent of the protection.  More work is necessary to reveal the effect of electric field poling and interfacial engineering on material stability.

\begingroup
\begin{table}
\caption{Reaction energetics (eV) of dissociated and oxidative water on PbI$_{2}$-terminated surfaces with different polarities}

\centering
\begin{tabular}{lccc}%
\hline
                                                                                                                                                                     &   \emph{P}$^{+}$  &   \emph{P}$^{-}$   \\ \hline
PbI$_{2}$$^{\rm surf}$ + H$_{2}$O $\rightarrow$ HO--PbI$_{2}$$^{\rm surf}$ + $\frac{1}{2}$H$_{2}$           &           1.46            &    1.52                    \\
PbI$_{2}$$^{\rm surf}$ + H$_{2}$O $\rightarrow$ O=PbI$_{2}$$^{\rm surf}$ + H$_{2}$                             &           3.02            &    4.35                     \\ \hline

\end{tabular}
\end{table}
\endgroup

In the interest of exploring beyond molecular adsorption of water, we constructed dissociated and oxidative water configurations on the PbI$_{2}$-terminated \emph{P}$^{+}$ and \emph{P}$^{-}$ surfaces.  The optimized structures can be seen in Figure S4.  The first case simulates H$_{2}$O dissociation by placing OH$^{-}$ and H$^{+}$ apart from each other.  The optimized structure for both \emph{P}$^{+}$ and \emph{P}$^{-}$, however, showed water convert from an initial dissociated state to a molecular form exhibiting similar bonding features to the previously described molecular adsorption configurations. Similarly to the structures described before, the oxygen was closer to Pb on the \emph{P}$^{-}$ than on the \emph{P}$^{+}$ surface.  We also studied a lone hydroxyl bonded to Pb to model oxidative adsorption.  The hydroxyl causes the Pb to rise up out of the surface.  Finally, we investigated a single oxygen atom for stronger oxidative adsorption.  On the \emph{P}$^{+}$ surface, the oxygen atom infiltrated the PbI$_{2}$ layer, bonding between the Pb and I.  On the \emph{P}$^{-}$ surface, the oxygen was incorporated into the surface plane, disrupting the periodicity of the PbI$_{2}$-terminated surface indicating the formation of PbI$_{2}$ units.  These oxidative adsorption configurations were higher in energy than molecular adsorption configurations (see Table 2), making them unlikely under low humidity conditions.  

\begingroup
\begin{table}
\caption{Elongation of top interlayer spacing, \emph{l}$_{\rm surf-subsurf}$, of PbI$_{2}$- and MAI-terminated surfaces and shortening of the vertical distance between H$_{2}$O and PbI$_{2}$-terminated surface, \emph{b}$_{\rm H_{2}O-Pb}$, with different polarities computed without and with polarizable continuum model (PCM) as an implicit solvation model for water.}

\centering
\begin{tabular}{lcccc}%
\hline
                                                                                                        &   Without PCM  &   PCM   &   Change after PCM is applied   \\ \hline
PbI$_{2}$-terminated \emph{P}$^{+}$, \emph{l}$_{\rm surf-subsurf}$ (\AA)            &           6.35         &    6.48   &     2.0 \%   \\
MAI-terminated \emph{P}$^{+}$, \emph{l}$_{\rm surf-subsurf}$ (\AA)                     &           6.83         &    7.08   &     3.7 \%    \\                         PbI$_{2}$-terminated \emph{P}$^{+}$, \emph{b}$_{\rm H_{2}O-Pb}$ (\AA)     &           2.68         &    2.58   &     -3.7 \%    \\
PbI$_{2}$-terminated \emph{P}$^{-}$, \emph{b}$_{\rm H_{2}O-Pb}$ (\AA)     &           2.57         &    2.50   &     -2.7 \%    \\ \hline

\end{tabular}
\end{table}
\endgroup

To approximate the effect of high humidity conditions on the MAPbI$_{3}$ surface, which is computationally expensive to treat with explicit water, we employ a polarizable continuum model (PCM) as the implicit solvation model on the MAI- and PbI$_{2}$-terminated bare \emph{P}$^{+}$ and \emph{P}$^{-}$ surfaces, as well as on the PbI$_{2}$-terminated \emph{P}$^{+}$ and \emph{P}$^{-}$ surfaces with one water molecule added explicitly.  These calculations were performed with the Joint Density Functional Theory package.~\cite{JDFTx,Arias92p1077,IsmailBeigi00p1} The bare surfaces show an elongation in the [001] (out-of-plane) lattice constant upon inclusion of solvation, as shown in Table 3. The PbI$_{2}$-terminated \emph{P}$^{+}$ and \emph{P}$^{-}$ surfaces with explicitly adsorbed water show a shorter O-Pb bond length, indicating a stronger bond.  The expansion normal to the surface when solvation is included suggests that higher water concentration weakens the interlayer bonding of the OMHP.  

\section*{\label{sec:level1} Conclusion}

In this work, we present an atomistic view of water interacting with MAPbI$_{3}$ (001) surfaces. We find that water favorably adsorbs on all the studied sites in both MAI- and PbI$_{2}$-terminated surfaces, supporting the existence of the experimentally observed hydrate state as a potential initial step of degradation of the material.  Our results do not show a direct reaction between the H$_{2}$O and MA molecule, which has been proposed to be an initial step of degradation.~\cite{Frost14p2584}  Although this is consistent with previous work,~\cite{Mosconi15p4885} we cannot completely rule out an acid-base reaction.  The specific bonding characteristics of water with the OMHP are termination and polarity dependent, with the MAI-terminated \emph{P}$^{+}$ surface having more favorable water adsorption than the corresponding \emph{P}$^{-}$ surface, while the PbI$_{2}$-terminated \emph{P}$^{-}$ surface binds water more strongly than the corresponding \emph{P}$^{+}$ surface.  These interactions arise from the hydrogen bond of the ammonium group of the MA molecule with the inorganic lattice and the water molecules, while the hydrophobic methyl group does not react with the water.  Harnessing the water-repelling nature of the methyl group of MA by poling the surface during device operation or creating a local domain through interface engineering could thus enhance the stability against degradation, although this stabilization strategy requires further investigation.

We demonstrate that physically introducing a water molecule below the top atomic layer of PbI$_{2}$-terminated surfaces pushes up the surface layer and affects the adjacent hollow sites,  providing important insight into how the material degradation could propagate.  The presence of more oxidizing adsorbates, though unlikely to form in low humidity conditions, has a greater effect on the surface structure.  Finally, calculations with an implicit solvation model of water provide a more general picture of water effects and show that the lattice elongates perpendicular to the surface in the presence of H$_{2}$O.  This implies that a higher water concentration may weaken the interlayer bonding strength of the hybrid perovskite, facilitating the dissociation of MAPbI$_{3}$ into its constituents.

\begin{acknowledgement}
N. Z. K. was supported by the US ONR under Grant N00014-12-1-0761, and by the Roy \& Diana Vagelos Scholars Program in the Molecular Life Sciences.
D.S.-G. was supported by the DOE under Grant DE-FG02-07ER15920.
F.W. was supported by the US ONR under Grant N00014-12-1-1033.
S.L. was supported by the NSF under Grant CBET-1159736.
A.M.R. was supported by the NSF under grant number CMMI-1334241.  Computational support was provided by the High-Performance Computing Modernization Office of the Department of Defense and the National Energy Research Scientific Computing Center.
\end{acknowledgement}

\section*{\label{sec:level1} Electronic Supplementary Information}
See Electronic Supplementary Information for projected density of states of structures reported and atomic structures of water adsorbed on MAPbI$_{3}$ in dissociated and oxidative configurations, $^{-}$OH and O$^{2-}$.

\bibliography{rappecites}

\end{document}